\documentstyle[preprint,revtex]{aps}
 
\begin{document}
 \preprint{}
 \draft

\begin{title}
The Collision of Two Black Holes
\end{title}

\author{Peter Anninos${}^{(1)}$, David
Hobill${}^{(1,2)}$, Edward Seidel${}^{(1)}$, Larry Smarr${}^{(1)}$
and Wai-Mo Suen${}^{(3)}$}

\begin{instit}
$^{(1)}$ National Center for Supercomputing Applications,
Beckman Institute, 405 N. Mathews Ave., Urbana, IL, 61801

$^{(2)}$ Department of Physics and Astronomy,
University of Calgary, Calgary, Alberta, Canada T2N 1N4

$^{(3)}$ McDonnell Center for the Space Sciences, Department of Physics,
Washington University, St. Louis, Missouri, 63130
\end{instit}

\receipt{}

\begin{abstract}
We study the head-on collision of two equal mass,
nonrotating black holes. We consider a range of cases
from holes surrounded by a common
horizon to holes initially separated
by about $20M$, where $M$ is the mass of each hole.
We determine the waveforms and energies radiated for both the
$\ell = 2$ and $\ell=4$ waves
resulting from the collision.  In all cases studied the
normal modes of the final black hole dominate the spectrum.
We also estimate analytically the total gravitational
radiation emitted,
taking into account the tidal heating of horizons using the
membrane paradigm, and other effects.
For the first time we are able to compare
analytic calculations, black hole perturbation theory, and strong
field, nonlinear numerical calculations for this problem, and we find
excellent agreement.
\end{abstract}



I. Introduction.  The collision of two black holes is considered to be one of
the most
promising and important astrophysical sources of detectable gravitational
radiation in our Universe~\cite{Abramovici}.
Since LIGO~\cite{Abramovici}
and VIRGO~\cite{Abramovici} are expected to begin taking data during this
decade, it is important to perform accurate calculations detailing the
shape and strength of the waveforms generated during such events.
The information gained from the detected
waveforms should allow one to reconstruct the astrophysical
parameters of the system, and will provide the first direct and
unambiguous evidence for the existence of black holes if the unique
signature of the quasinormal modes~\cite{Chandra} of the hole is excited.

In this paper
we present results of numerical studies of time symmetric, axisymmetric
head-on collisions of equal mass black holes.  We have
been able to extract the waveforms and the total energy emission
resulting from the collision.  Analysis of the waveforms
reveals clearly for the first time that the quasinormal modes
of the final black hole are strongly excited from the collision.
This work extends and
refines
the early work of Smarr and Eppley~\cite{smarrphd,smarrsources} that
suggested that the
normal modes of the final hole were excited, but the resolution and
waveform extraction techniques available at that time did not
permit a clean and unambiguous matching to black hole normal modes.
The numerical
difficulties inherent in this problem also led to fairly
large uncertainties in the total
energy radiated (Smarr quotes a probable uncertainty factor of
2~\cite{smarrsources}).
Because of the importance of this
fundamental physical problem, we have revisited this calculation with the
benefit of more powerful computers and improved analytic and
numerical techniques developed over the intervening 15 years to calculate
unambiguous waveforms and energy fluxes resulting from the collision.

The initial data sets that we adopt are analytic solutions
originally discovered by
Misner~\cite{Misner}
and subsequently analyzed and evolved
by Smarr and Eppley.
They are characterized by a parameter $\mu$
determining the mass $M$ of a hole
and the proper separation $L/M$ of the
two holes (see Table 1), and consist of two throats connecting
identical asymptotically flat spacetimes
In this paper we apply the code described in
Ref.~\cite{Anninos93x}
to compute evolutions for a
family of Misner spacetimes representing equal mass black holes colliding from
distances of between $4M$ and $20M$.
If the throats are close enough together
(for $\mu < 1.36$)~\cite{Smarr76}
a common apparent horizon surrounds them both,
so that the system really represents a single, highly perturbed
black hole.  For throats separated by more than about $8M$
we are confident that there is no common event horizon surrounding them,
as shown by directly integrating light rays (see also a hoop
conjecture argument~\cite{Anninos93y}).

II. Waveforms.  Numerically generated waveform templates
will be essential for analysis of data collected by future gravitational
wave detectors~\cite{Abramovici}.
The method we use to calculate waveforms
is based on the gauge invariant extraction technique
developed by Abrahams and
Evans~\cite{AE88} and applied in Ref.~\cite{ABHSS92} to black hole
spacetimes.

For all of the cases studied in this paper we have
extracted both the $\ell = 2$ and $\ell = 4$ waveforms at radii of
$30, 40, 50, 60,$ and $70M$.
By comparing results
at each of these radii we are able to check the propagation of waves
and the consistency of our energy calculations.  In general we find
agreement among detectors at different radii to be
within 10-20\%.
We have also run
all the calculations with resolutions of 200x35 (radial,angular)
and 300x55 with only minor differences in
the $\ell=2$ waveforms and the total energy radiated.  (But as detailed
in Ref.~\cite{Anninos93y} the amplitude and precursor of the  $\ell=4$ waveform
are rather sensitive
to computational parameters.)

In Fig. 1 we show the $\ell=2$ extracted waveform for the case $\mu=2.2$,
for which there is no initial common event horizon.
The solid line shows the waveform detected at a distance $r=40M$
and the
long dashed line shows the waveform extracted at $r=60M$. The
wave is clearly propagating away from the hole at light speed with
essentially invariant shape and amplitude, with a wavelength of $2\times16.8M$,
confirming the original findings of Smarr and Eppley~\cite{smarrsources}.
However, our more accurate code now allows us to go beyond
estimating the wavelength and to fit quantitatively the waveform
to results known from black hole perturbation theory.
The short dashed line shows
the result of fitting the $r=40M$ waveform (in the range
$64M<t<160M$)
to a linear combination of the
fundamental and first excited state of the $\ell=2$ quasinormal mode for
the final black hole with mass $2M$.
The fit is quite good, matching both the wavelength and damping time,
showing that the final
black hole mass is indeed very close to the total mass of the spacetime.

In Fig. 2 we show the $\ell=4$ waveform for the same case, extracted at the
same radius.
Again this waveform has been fit (over
a similar range) to a superposition of the fundamental and first
excited $\ell=4$ quasinormal modes of the black hole.  Waveforms
for all cases studied show similar behavior:  the normal modes of the
final black hole are excited and account for most of the emitted signal.

III.  Energy Output.
We first present
a semi-analytic determination of the energy radiated by the
colliding black holes. Such an estimate is interesting not just as a
confirmation of
the numerical results, but more importantly, it provides physical understanding
of the numerical data.

III.A.  Analytic Estimate.  We base our understanding of the two black
hole collision on the well-studied
problem~\cite{Davis71,Davisetal} of a test
point particle $m$ at rest from infinity plunging into
a Schwarzschild black hole $M$. The gravitational wave energy radiated for the
test
particle case is found to be~\cite{Davis71}
\begin{equation}
\label{pertenergy}
E = 0.0104 {m^2 / M} ~~~, for ~ m \ll M.
\end{equation}
We build on this accurate result to approximate the energy
loss for the case of two unequal mass black holes of mass $m$ and $M$,
with $m \leq M$.  We include correction factors into Eq.
(\ref{pertenergy}), so that it can
better approximate the two black hole collision.
The most important effects
one has to take into account~\cite{Anninos93y} are that, ({\it i})~$m$
may be comparable to $M$, ({\it ii})~the infall is not from
infinity, ({\it iii})~the black hole, unlike a point particle, has internal
dynamics.

It is useful to first understand the $m^2/M$ dependence of
Eq.~(\ref{pertenergy}).
In Newtonian approximations,  the quadrupole moment of the system is
$I \sim m r^2$ for $m \ll M$,
where $r$ is the radial distance between $m$ and $M$. The gravitational wave
luminosity $L$ is given by
\begin{equation}
\label{lum}
L \propto \overdots{I}^2
\sim m^2 \bigl(\ddot{r}\dot{r}\bigr)^2
\sim m^2 \biggl( {M^3 / r^5}\biggr) \quad . \end{equation}
As most of the radiated energy is emitted when $m$ is falling near the horizon
of $M$, to obtain the total energy radiated $E$, we can evaluate the
luminosity at $r \sim M$ and multiply it by a time scale of $M$.
This gives $E \propto {m^2 / M}$ as in Eq. (\ref{pertenergy}).
Now for $m$ not much smaller than $M$, it is
more accurate to use $I \sim \mu r^2$,
with $\mu \equiv (mM)/(M+m)$.
The quadrupole moment formula then leads to~\cite{smarrsources} $ E \propto
\mu^2/M$, and we modify Eq.~(\ref{pertenergy}) to be
\begin{equation}
\label{energyformula}
E = 0.0104 {\mu^2 / M}.
\end{equation}
Notice that for $m=M$, the $\mu^2$ of
Eq.~(\ref{energyformula}) introduces
a quite significant factor of $1/4$.

The quadrupole formula (2) also suggests how the expression should be modified
when the infall is not from infinity.
There are two effects: ($i$)~there is less time to radiate
when falling from a finite distance, and ($ii$)~the infalling velocity is
smaller for the same separation. The latter effect is much more important.
These two effects can be combined into a single factor $F_{r_o},$
\begin{equation}
\label{fro}
F_{r_o}
= {E_{r_o} / E_{\infty}}
= {\int_{r_o}^{2M} \dot{r}(\ddot{r})^2 dr /
\int_{\infty}^{2M} \dot{r}(\ddot{r})^2 dr } .
\end{equation}
Here r is the radial Schwarzschild coordinate, $r_o$ is
the starting point of the infall, and for the infall velocity $\dot{r}$ we
use the relativistic expression
$ ( 1- {2M / r} ) \sqrt{{2M / r}-{2M / r_o}}
/ \sqrt{{1-2M / r_o}}$~\cite{qnote}.  We find $F_{r_o}$ ranges from $\sim
0.4$ (for $r_o = 6 M$) to 1 (for $r_o = \infty$).

All considerations up to this point are the same whether the infalling
object is a
point particle or a black hole.
As far as the gravitational wave output is concerned, the most important
difference between a point mass and a black hole is that a black hole has
internal dynamics.
There are more channels into which the initial gravitational potential energy
in the system can dissipate.
Such dissipation decreases the kinetic energy
and hence the velocity of the infalling hole, reducing the wave output.

When the hole $m$
is falling in the static gravitational field generated by $M$, the horizon of
$m$
will be deformed by the tidal force. In the membrane paradigm~\cite{membrane}
of a black
hole, in which the horizon is treated as a 2-D surface living in a 3-D space,
endowed with physical properties such as viscosity, this tidal deformation
heats up
the horizon. The heating is described by the horizon
equations~\cite{membrane,Suen86}
$- \dot{\sigma_{ab}}
+ (g- \theta ) \sigma_{ab}
+ \bigl( 2\sigma_{ac} + \gamma_{ac} \theta \bigr)\sigma_b^c = \epsilon_{ab}$,
$- \dot{\theta} + g\theta
- \theta^2/2 =\sigma_{ab}\sigma^{ab}$, and
$\dot{\gamma_{ab}}
= 2\sigma_{ab} + \gamma_{ab}\theta$.
Here $\gamma_{ab}$ is the 2-D metric of the horizon,
$\theta$ and $\sigma_{ab}$ are the expansion rate and shear of the
horizon generators,and $g=1/(4m)$ is the surface gravity, while
$\epsilon_{ab}$ is the normalized electric
part of the Weyl tensor [$C_{a\mu b\nu} l^\mu l^\nu$, with $l^\mu$ the
horizon generators], here to be evaluated on the horizon of the infalling
hole $m$ as it falls
in the external tidal field ($M / r ^ 3 $ ) of the hole $M$.
For our present purpose, it suffices to solve these equations in the linear
approximation, dropping all terms quadratic in the horizon deformation.
The Green's function for the linearized equations satisfying the teleological
boundary condition is simply $G(t,t^\prime ) =\exp [g(t-t^\prime )]$ for
$t<t^\prime$ and  $G(t,t^\prime ) =0$ for $t>t^\prime$.
Hence $\theta$ can be obtained by a simple integration.
The fraction of the available
gravitational potential energy originally in the system which is dissipated
into the heating of the horizon of the infalling hole is given by
\begin{eqnarray}
\label{fh}
{\Delta m \over m}
= {1\over 2} \int \theta \, dt = {1\over 2} \int_{r_o}^{2m+2M}
[\theta / \dot{r} ] dr. \end{eqnarray}
The integration in Eq. \ref{fh} is terminated at the point when the two holes
are engulfed by a common horizon.
For $m=M$, the
heating on the horizon of $M$ is the same as that on $m$, so the total fraction
of energy going into horizon heating
is given by two times Eq. \ref{fh}, i.e.,
the reduction factor for the energy available for wave generation is
$F_h =1-2 {\Delta m / m }$, which
decreases with increasing initial separation.
For the range of
$r_o$ studied here, $F_h$ gives a reduction
from about 3 to 13\%.

With these correction factors taken into account, the total gravitational
energy radiated is given by
\begin{equation}
\label{totalenergy}
E = 0.0104 {\mu^2 / M} \times F_{r_o} \times F_h .
\end{equation}
There are other correction factors in Eq.~(\ref{totalenergy}) that we studied
and included in Ref.~\cite{Anninos93y}.
As their effects are much smaller, we shall not
discuss them here.  We note that for holes initially infinitely separated,
$F_{r_o} $ is 1, while $F_h$ is 0.86.
In the next subsection we compare
Eq.~(\ref{totalenergy}) for the total energy radiated
to numerical results.

III. B. Numerical Results.
The waveforms
shown in Figs. 1,2 have been normalized so that the total
energy carried away from the system is given by
$\dot{E} =\dot{\psi}^2/(32\pi)$
for all $\ell$ modes, where $\psi$ is the extracted waveform.  The total energy
$E$ is computed from this expression.
For a check on the accuracy and consistency of
this calculation, we have computed $E$ at all five detectors
listed above, and at both high and medium grid
resolution.  We have also computed the energy loss via two
independent curvature based indicators, the Newman-Penrose
Weyl tensor component $\Psi_4$ and the Bel-Robinson poynting vector
These constructions yield results
consistent with the
gauge invariant formulations when computed
in the asymptotic far field spacetime~\cite{Anninos93y}.

In Fig.~3, we compare results of various analytic estimates of the energy loss
to our numerical results for
the dominant $\ell=2$ waveform calculation.
The connected circles show the maximum possible radiation output obtained by
comparing the initial black hole masses estimated by the areas of the
horizons (or a single horizon if the holes
are close enough) to the total mass of the spacetime.
For large separation this number approaches 29\% as expected from the
work of Hawking~\cite{Hawking}.  The six clusters of unconnected symbols
result from our numerical simulations for the six cases studied (listed
in Table I).  Each symbol corresponds to the $\ell=2$ energy
computed at each of five ``wave detectors'' located throughout the
radiation zone as before.

The analytic estimate for total gravitational
wave energy output given by Eq.~\ref{totalenergy}
is plotted as a dashed line.
The agreement of the analytic and the numerical result is highly remarkable,
as the analytic and numerical results were obtained in a double blind
manner.  For $L/M$ less than about 9 the
analytic formula overestimates the actual energy output computed
numerically.
This is to be expected since for small enough separation the holes are
initially engulfed by a common event horizon and the approximation
for colliding black holes is inappropriate.  For reference, the early
results of Smarr and Eppley are plotted as large crosses with error
bars suggested by Smarr~\cite{smarrsources}.  Within the large errors
quoted, those early results are remarkably consistent with our results.

III. Conclusions.
We have performed numerical and analytic calculations predicting the
gravitational waveforms generated and total gravitational wave energy
emitted when two equal mass black holes
collide head on.  Both the $\ell=2$ and $\ell=4$ waveforms are fit
remarkably well by a superposition of the fundamental and first excited
state of the black hole quasinormal modes, and the total energy output
is found to be on the order of $0.002M$.
The analytic study
confirms and elucidates the
numerical results, matching the numerical results remarkably well.
The total energy output
is far below estimates given by a simple application of the area theorem.
Taken together, the analytic and numerical results
indicate that even for holes that are initially infinitely separated,
the total energy output will be the same order of magnitude.

We would like to thank David Bernstein for a number of
helpful discussions, Joe Libson for integrating light rays
through the spacetimes, and Mark Bajuk for his work on
visualizations of our numerical simulations that aided greatly
in their interpretation.  This work was supported by NCSA, NSF
Grant~91-16682, and NSERC Grant No. OGP-121857, and calculations
were performed at NCSA and the Pittsburgh Supercomputing Center.

\figure{$\ell = 2$ waveforms for the case $\mu=2.2$.  The solid line shows
the waveform extracted at $R=40M$ and the long dashed line shows the waveform
at $R=60M$.  The short dashed line shows the quasinormal mode fit.
\label{fig1a}}
\figure{
$\ell = 4$ waveforms for the case $\mu=2.2$.  The solid line shows
the waveform extracted at $R=40M$
and the short dashed line shows the quasinormal mode fit.
\label{fig1b}}
\figure{
The total gravitational wave energy output $E$
shown for various cases. The connected circles are the upper limit based
on the area theorem, the clustered symbols show numerical results at various
detector locations, the dashed line is the semi-analytic estimate, and the
crosses show early results by Smarr and Eppley with
their approximate error bars.
\label{fig2}}
\begin{table}
\begin{tabular}{|c|c|c|c|} \hline
 $\mu$     & $M$ & $L/M$ & Apparent horizon\\ \hline
1.2  & 1.85  & 4.46 & global  \\ \hline
1.8  & 0.81  & 6.76 & separate  \\ \hline
2.2  & 0.50  & 8.92 & separate  \\ \hline
2.7  & 0.29  & 12.7 & separate  \\ \hline
3.0  & 0.21  & 15.8 & separate  \\ \hline
3.25  & 0.16  & 19.1 & separate  \\ \hline
\end{tabular}
\caption{The physical parameters of the six initial data sets
studied are summarized.
$M$ is the mass parameter defined in the text,
$L/M$ is the proper distance between the throats, and we note whether or not
one
apparent horizon surrounds both holes.
\label{Table}}
\end{table}

\begin{references}
\bibitem{Mass}
For two throats not well separated we define $M=M_{adm}/2$ ($M_{adm}$
is the total spacetime mass).
\bibitem{qnote}
This expression for $F_{r_o}$ represents one (non-unique) way of extending the
quadrupole formula to the relativistic regime.
This choice is convenient for our present purpose.
\end{references}
\end{document}